\documentclass{jetpl}
\usepackage{graphicx}

\twocolumn \lat

\begin{document}
\title{Analog of Fishtail Anomaly in Plastically Deformed Graphene}
\rtitle{} \sodtitle{Analog of Fishtail Anomaly in Plastically
Deformed Graphene}
\author{S. Sergeenkov and F.M. Araujo-Moreira}
\address{Grupo de Materiais e Dispositivos,
Departamento de F\'{i}sica, Universidade Federal de S\~{a}o
Carlos, 13565-905 S\~{a}o Carlos, SP, Brazil }

\date{\today}

\abstract{ By introducing a strain rate $\dot {\epsilon}$
generated pseudo-electric field $E^d_x\propto \hbar \dot
{\epsilon}$, we discuss a  magnetic response of a plastically
deformed graphene. Our results demonstrate the appearance of
dislocation induced paramagnetic moment in a zero applied magnetic
field. More interestingly, it is shown that in the presence of the
magnetoplastic effect, the resulting magnetization exhibits
typical features of the so-called fishtail anomaly. The estimates
of the model parameters suggest quite an optimistic possibility to
experimentally realize the predicted phenomena in plastically
deformed graphene.}


\maketitle

{\bf 1. Introduction.} Many interesting and unusual phenomena due
to significant modifications of the carbon based materials
(including graphene) under mechanical deformations leading to
generation of strong intrinsic pseudomagnetic fields have been
recently discussed (see, e.g.~\cite{1,2,3,4,5}  and further
references therein). Of special interest for us are dislocations
related properties in plastically deformed graphene and carbon
nanotubes~\cite{6,7,8,9}.

In this Letter we consider theoretically some intriguing magnetic
properties of graphene sheet under plastic deformations by
directly incorporating a constant strain rate as a time dependent
gauge potential into the Dirac model. The physics behind our
findings and feasibility of their experimental verification are
discussed.

{\bf 2. Model.} Recall~\cite{1,2,5} that in the absence of
chirality (intervalley) mixing, the low-energy electronic
properties of graphene near the Fermi surface can be reasonably
described by a two-component wave function $|\Psi>=(\Psi_1,
\Psi_2)$ obeying a massless Dirac equation
\begin{equation}
i\hbar \frac{\partial |\Psi>}{\partial t}= {\cal H}|\Psi>
\end{equation}
with an effective Hamiltonian
\begin{equation}
{\cal H}=v_F(\sigma_x \pi_x+\sigma_y \pi_y)
\end{equation}
Here, $\pi_a=p_a+eA_a+eA^d_a$ with $p_a=-i\hbar \nabla_a$ being
the momentum operator, $A_a$ the electromagnetic vector potential,
and $A^d_a$ the deformation induced vector potential; $\sigma_a$
are the Pauli matrices, and $v_F$ is the Fermi velocity. In what
follows, $a =\{x,y\}$. Let us consider a graphene sheet of length
$L$ and width $W$ under the simultaneous influence of plastic
deformation and perpendicular applied magnetic field $B_z$
(defined via the vector potential $A_y=B_zx$). As is well
known~\cite{1,2,5}, the homogeneous strain $\epsilon$ induced
gauge potential, given by $eA^d_x=\hbar \epsilon /r$ (where
$r=0.14nm$ is carbon-carbon bond length), leads to appearance of
intrinsic pseudo-magnetic field $B^d_z=\hbar {\epsilon} /eLr$
inside deformed graphene lattice. By analogy with an applied
electric field $E_x$ (defined via time-dependent vector potential
$A_x=E_xt$), we introduce plastic deformation effects into the
model through a constant plastic strain rate $\dot {\epsilon}$
dependent vector potential $A^d_x=E^d_xt$ resulting in appearance
of intrinsic pseudo-electric field $E^d_x=\hbar \dot {\epsilon}
/er$.

\begin{figure}
\centerline{\includegraphics[width=7.0cm]{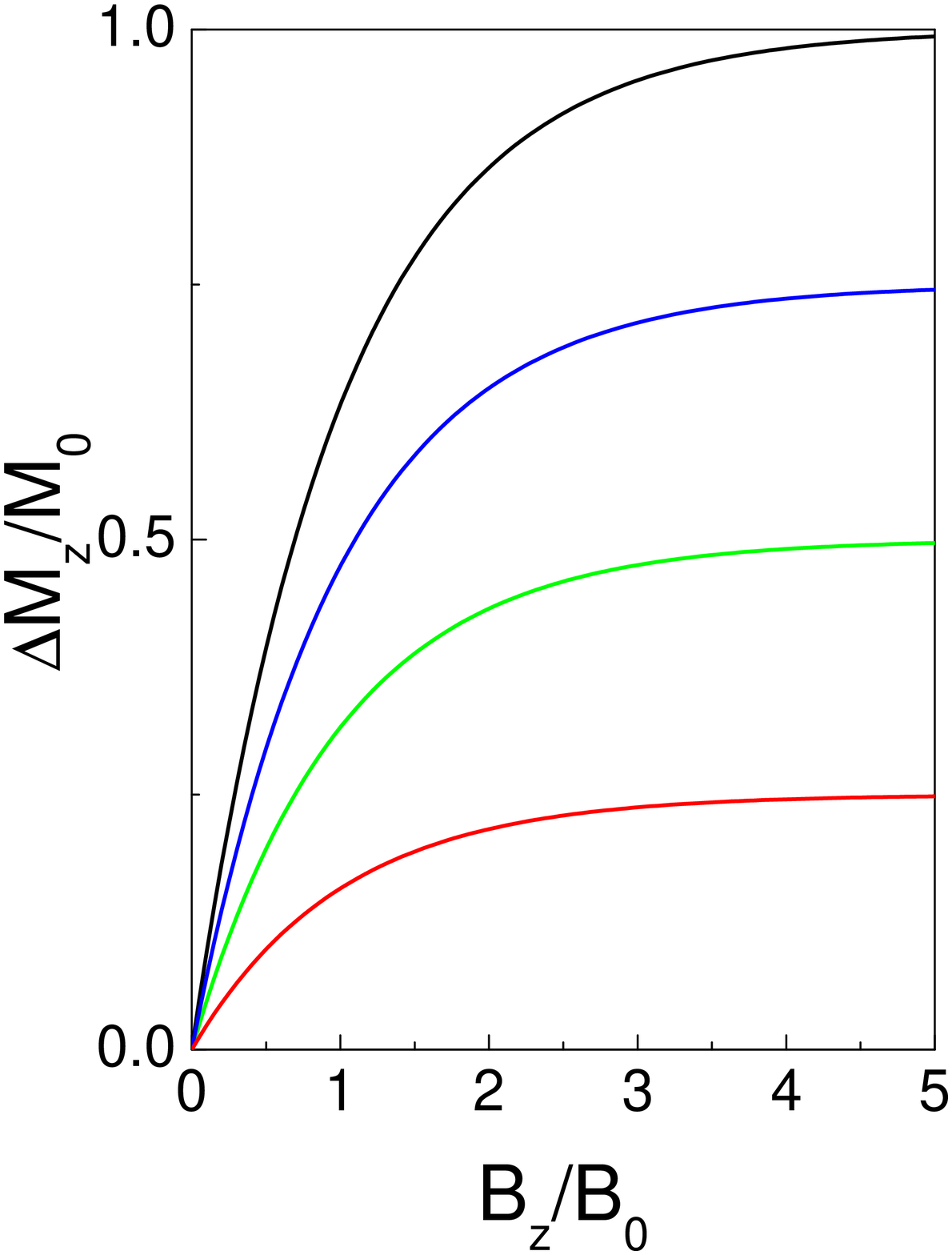}}\vspace{0.25cm}
\caption{Fig.\ref{fig:fig1}. The magnetic field dependence of the
normalized magnetization in plastically deformed graphene for
different values of the normalized strain rate (from bottom to
top): $\dot {\epsilon}/\dot {\epsilon}_0=0.25, 0.5, 0.75,$ and
$1.0$. Here, $\dot {\epsilon}_0=10^{-4}s^{-1}$.} \label{fig:fig1}
\end{figure}

\begin{figure}
\centerline{\includegraphics[width=7.0cm]{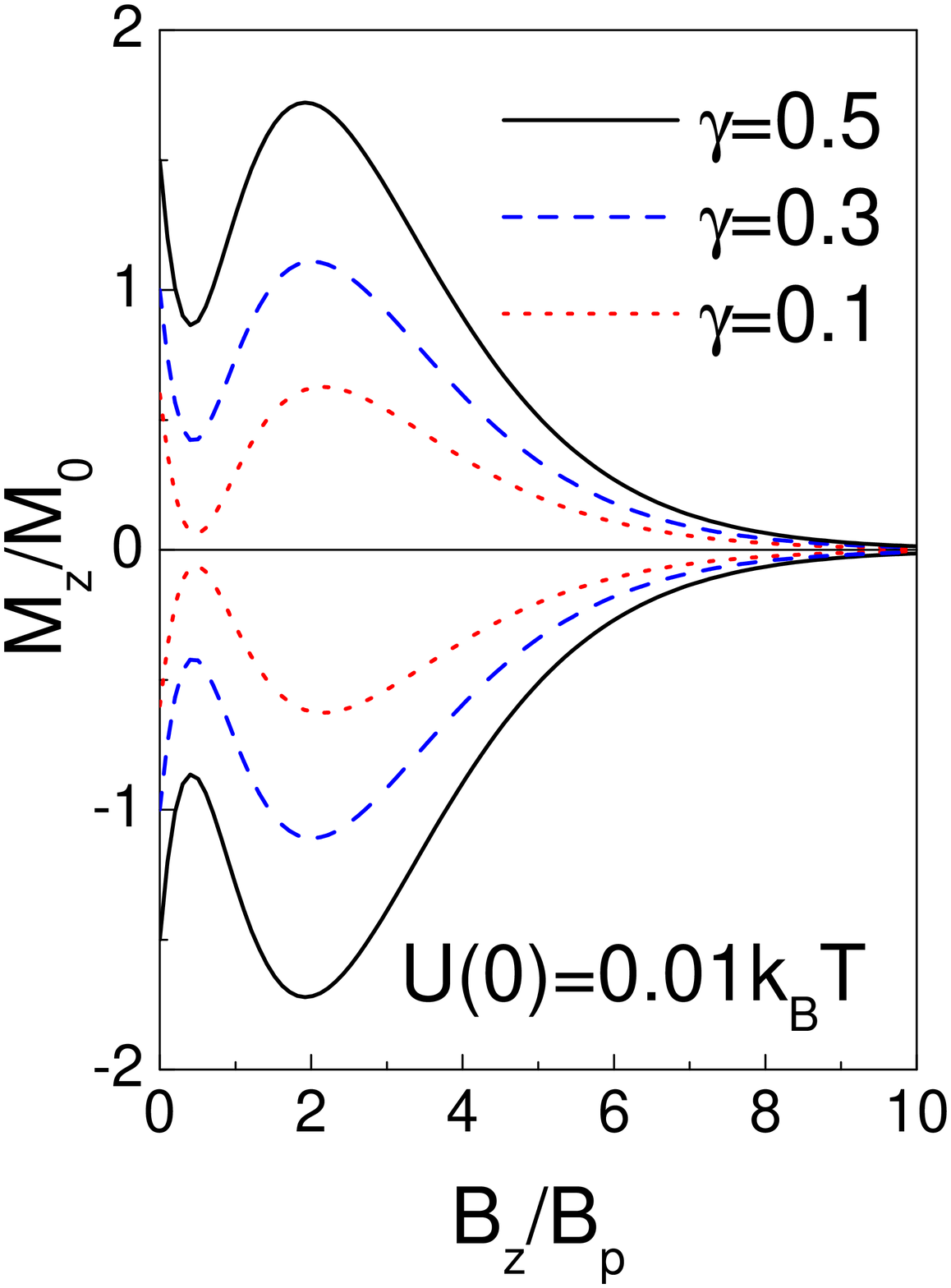}}\vspace{0.25cm}
\caption{Fig.\ref{fig:fig2}. The magnetic field dependence of the
normalized magnetization in the presence of magnetoplastic effect
for $U(0)=0.01k_BT$, $\dot {\epsilon}(0)=10^{-4}s^{-1}$ and three
values of $\gamma=B_p/B_0$. A "diamagnetic" part of the curve
$-M_z/M_0$ is added for better visual effects only.}
\label{fig:fig2}
\end{figure}

{\bf 3. Results and Discussion.} Let us consider the magnetic
response of the graphene sheet (with area $S=LW$) on plastic
deformation by analyzing its magnetization:
\begin{equation}
M_z(B_z, \dot {\epsilon}) \equiv -\frac{1}{S}\left [\frac{\partial
{\cal E}(B_z, \dot {\epsilon})}{\partial B_z}\right ]
\end{equation}
Here
\begin{equation}
{\cal E}(B_z, \dot {\epsilon})=\int_0^\tau \frac{dt}{\tau}
\int_0^L\frac{dx}{L}
\int_{0}^W\frac{dy}{W}\sum_{i=1}^2\left<\Psi_i|{\cal
H}|\Psi_i\right>
\end{equation}
is the total energy of the problem based on the previously
obtained~\cite{5} solutions $|\Psi_i>$ of time-dependent Eq.(1)
($\tau$ is the characteristic time related to duration of plastic
deformation, that is $\dot {\epsilon}\simeq \epsilon/\tau$). First
of all, the analysis of Eqs.(1)-(4) reveals that plastic
deformation results in appearance of a non-zero magnetic moment
$\mu_z=M_z(0, \dot {\epsilon})S=\hbar v_F\tau \dot {\epsilon}/r$
in a zero applied magnetic field ($B_z=0$). For typical
experimental values of the applied strain rates~\cite{6,9} $\dot
{\epsilon}\simeq 10^{-4}s^{-1}$, we obtain $\mu_z \simeq 1\mu_B$
for a reasonable estimate of the plastically induced paramagnetic
moment in graphene ~\cite{10} ($\mu_B$ is the Bohr magneton).
Fig.~\ref{fig:fig1} shows the field dependence of the induced
magnetization $\Delta M_z=M_z(B_z, \dot {\epsilon})-M_z(0, \dot
{\epsilon})$ for different values of the normalized strain rate
$\dot {\epsilon}$ where $B_0=\Phi_0/S$ is a characteristic
magnetic field ($\Phi_0$ is the flux quantum). It is worth noting
that it closely follows the observed~\cite{10} behavior of the
point defects induced paramagnetic moment in graphene for
different values of density of vacancies $\rho_v$. This makes
sense because plastic deformation is driven by motion of
dislocations (with velocity $v_d$) leading to strain rate
dependence on both $v_d$ and dislocation density $\rho_d$ as
follows, $\dot {\epsilon}=b\rho_dv_d$. Here $b$ is the absolute
value of the relevant Burgers vector. It is instructive to point
out that the dislocation velocity $v_d$ in a sense plays a role of
the Fermi velocity $v_F$ (which links applied electric and
magnetic fields as $E=v_FB$) in relationship between strain rate
induced pseudo-electric $E^d$ and strain induced pseudo-magnetic
$B^d$ fields. Indeed, with quite a good accuracy we can write
$E^d\simeq v_dB^d$. At the same time, it is important to emphasize
that (in addition to a definitely non-universal character of
$v_d$) these two characteristic velocities describe phenomena on a
completely different scale because while $v_F\simeq 10^{6}m/s$,
typical dislocation velocities rarely exceed $v_d\simeq
10^{-6}m/s$.
Let us turn now to another interesting phenomenon related to
behavior of plastic deformation under applied magnetic field.
Namely, as it was experimentally observed for many different types
of materials (including semiconductors), upon application of
magnetic field (of the order of $B=1T$), dislocation velocity
$v_d$ (and hence strain rate) becomes strongly field dependent.
This phenomenon is called a magnetoplastic effect (MPE). It was
discovered by Al'shits et al in 1987~\cite{11}. There are many
different mechanisms which could be responsible for such a
behavior~\cite{12}. One of them (and probably most appropriate for
graphene~\cite{10}) is based on interaction between uncompensated
spin of dislocation's core and point paramagnetic
impurities~\cite{13,14} due to the difference in gyromagnetic
factors $g$ (so-called $\Delta g$ mechanism) leading to appearance
of resonance frequency $\omega_r=\Delta g \mu_B B/\hbar$ in
applied magnetic field $B$. As a result, the interaction energy
$U$ between dislocation and impurity becomes field dependent with
$U(B)>U(0)$, which in turn leads to a significant increase of the
thermally activated dislocation velocity described by the
following expression~\cite{13,14}
\begin{equation}
v_d(B)=v_d(0)\exp  \left[ \frac{\Delta U(B)}{k_BT} \right ]
\end{equation}
where
\begin{equation}
\Delta U(B)=U(B)-U(0)=U(0)f(B)
\end{equation}
with
\begin{equation}
f(B)=\frac{B^2}{B^2+B_p^2}
\end{equation}
Here, $B_p=\hbar/\Delta g \mu_B\tau_s$ is the characteristic field
for manifestation of MPE with $\tau_s\simeq \omega_r^{-1}(B=B_p)$
being the characteristic time. For typical values of $\Delta
g\simeq 10^{-3}$ and $\tau_s \simeq 10^{-8}s$, we get $B_p\simeq
1T$ for the estimate of the intrinsic magnetic field due to
spin-mediated interaction between point and linear
defects~\cite{11,12,13}. So far, we have ignored the MPE in the
magnetic response of graphene under plastical deformation. Let us
see now what happens with magnetization in the presence of the
above discussed MPE, that is assume that the strain rate becomes
field dependent as follows, $\dot {\epsilon}(B_z)=b\rho_dv_d(B_z)$
with $v_d(B)$ given by Eq.(5). Notice that accounting for MPE
virtually transforms our pseudo-electric field $E^d_x$ into a
pseudo-magnetoelectric one $E^d_x(B_z)$. The obtained magnetic
field dependence of the resulting magnetization is shown in
Fig.~\ref{fig:fig2} for $U(0)=0.01k_BT$, $\dot
{\epsilon}(0)=10^{-4}s^{-1}$ and for different values of the ratio
$\gamma=B_p/B_0$ between two characteristic fields (notice that in
Fig.~\ref{fig:fig2} the applied field is normalized to $B_p$
instead of $B_0$ as in Fig.~\ref{fig:fig1}). We observe a
remarkable fishtail like behavior of magnetization in plastically
deformed graphene in the presence of MPE (a "diamagnetic" part of
the curve $-M_z/M_0$ in Fig.~\ref{fig:fig2} is added for better
visual effects only). As it is clearly seen, the curve first
reaches minimum at $B_z/B_p\simeq 1$, peaks around $B_z/B_p\simeq
2$ and then gradually diminishes at higher applied fields. Recall
that such a behavior has been observed before in superconductors
and attributed to a perfect match between the sizes of the vortex
core and the pinning center. While for Abrikosov vortices the best
pins are point defects (vacancies)~\cite{15}, the so-called
Josephson vortices (fluxons) require linear (or even planar)
defects for their effective pinning~\cite{16}. By analogy, we can
assume that the discussed here MPE induced fishtail anomaly in
graphene structure probably has something to do with a perfect
match (energy minimization) between a paramagnetic impurity and
magnetic field modified dislocation, which serves as a
spin-sensitive pinning center for this impurity~\cite{7,8,10}.
For typical values of the width $W=40nm$ and aspect ratio $L/W=10$
we obtain $B_0=\Phi_0/WL \simeq 10T$ for the estimate  of the
characteristic field in graphene (shown in Fig.~\ref{fig:fig1})
which should be compared with the earlier estimated value of the
MPE mediated intrinsic magnetic field $B_p\simeq 1T$. According to
Fig.~\ref{fig:fig2}, the fishtail like behavior is expected to
manifest itself already for $\gamma=B_p/B_0\ge 0.1$ which makes
the experimental observation of the predicted here phenomena quite
feasible.

In summary, by incorporating strain rate as a gauge potential (for
pseudo-electric fields) into the model describing low-energy
electron properties in graphene, we have calculated magnetic
response of a plastically deformed graphene. According to our
findings, under such deformations graphene acquires dislocation
induced paramagnetic moment in a zero applied magnetic field.
Besides, in the presence of the so-called magnetoplastic effect
(when the strain rate becomes strongly dependent on the applied
magnetic field), the resulting magnetization was found to exhibit
typical features of the so-called fishtail anomaly, attributed to
spin-dependent interaction between dislocations and paramagnetic
impurities in plastically deformed graphene.

This work has been financially supported by the Brazilian agencies
CAPES, CNPq, and FAPESP.

\end{document}